# Wave modes of collective vortex gyration in dipolar-coupled-dot-array magnonic crystals

Dong-Soo Han[1], Andreas Vogel[2], Hyunsung Jung[1], Ki-Suk Lee[1†], Markus Weigand[3], Hermann Stoll[3], Gisela Schütz[3], Peter Fischer[4], Guido Meier[2] and Sang-Koog Kim[1*]

1. National Creative Research Initiative Center for Spin Dynamics and Spin-Wave Devices, Nanospinics Laboratory, and Research Institute of Advanced Materials, Department of Materials Science and Engineering, Seoul National University, Seoul 151-744, Republic of Korea

2. Institut für Angewandte Physik und Zentrum für Mikrostrukturforschung, Universität Hamburg, 20355 Hamburg, Germany

3. Max-Planck-Institut für Intelligente Systeme, 70569 Stuttgart, Germany

4. Center for X-ray Optics, Lawrence Berkeley National Laboratory, Berkeley CA 94720, USA

*Correspondence and requests for materials should be addressed to Sang-Koog Kim (email: sangkoog@snu.ac.kr).

†Current address : School of Mechanical and Advanced Materials Engineering, Ulsan National Institute of Science and Technology, Korea

Lattice vibration modes are collective excitations in periodic arrays of atoms or molecules. These modes determine novel transport properties in solid crystals. Analogously, in periodical arrangements of magnetic vortex-state disks, collective vortex motions have been theoretically predicted. Here, we experimentally observe discrete wave modes of collective vortex gyration in a one-dimensional (1D) chain of periodic disks using time-resolved scanning transmission x-ray microscopy. The observed discrete modes are interpreted based on micromagnetic simulation and numerical calculation of coupled linearized Thiele equations. Dispersion of the modes in the first Brillouin zone is found to be strongly affected by both vortex polarization and chirality ordering, as revealed by the explicit analytical form of the 1D infinite chain. A thorough understanding thereof is fundamental both for lattice vibrations and vortex dynamics, which we demonstrate for 1D magnonic crystals. Such magnetic disk arrays with vortex-state ordering, referred to as magnetic metastructure, offer potential implementation into information processing devices.

Recently, collective spin excitations in nano-scale magnetic elements, particularly spin waves, have become a focus of attention in nanomagnetism and related spintronics, owing to their potential implementation in information processing devices[1-12], in addition to the advances made in the understanding of fundamental modes in such geometrically confined spin systems[13]. New advances in both nanofabrication technology[14] and time- and space-resolved measurement techniques[1,2,15] have enabled intensive studies of a wide variety of magnonic crystals (MCs) such as one-dimensional (1D) strips[16-20], two-dimensional (2D) arrays of magnetic nanoelements[21-24], and antidot lattices of periodic holes having a circular or rectangular shape in 2D continuous films[25-27]. Furthermore, technological interest in the practical applicability of MCs to future information storage and processing devices[28-29] is rapidly growing. In patterned MCs, band structures including band widths and gaps can, in principle, be tailored through their constituent materials and the isolated elements' dimension and separation distances[1-3,16-25]. However, despite recent insight into the allowed magnonic modes in a rich variety of MCs, collective vortex-gyration modes in vortex-state arrays remain elusive, notwithstanding Shibata *et al.*[30-31]'s theoretical prediction of dipolar-coupled vortices in 2D magnetic disk arrays and the experimental demonstrations of vortex-gyration transfer between two (or more) coupled disks[32-43].

Here, we report on the first direct experimental demonstration, by means of state-of-the-art time-resolved scanning transmission x-ray microscopy (STXM), of quantized (or discrete) wave modes of collective vortex gyrations excited in five physically separated but dipolar-coupled disks in a permalloy (Py: $Ni_{80}Fe_{20}$) disk chain. With the help of numerical calculation, micromagnetic simulations, and analytical derivations, we investigate the experimentally observed discrete modes and their dispersion relations. The results reveal that characteristic dispersions are expressed in terms of the intrinsic angular eigenfrequency $\omega_0$ of isolated disks and the specific polarization $p$ and chirality $C$ ordering. The underlying physics can be well understood in terms of the dynamic dipolar interaction associated with the specific $p$ and $C$ orderings. Accordingly, and promisingly, the propagation property of collective vortex gyration and its dispersion can be manipulated by vortex-state ordering, the dimensions of each disk, and the nearest-neighbouring (NN) disks's interdistance. This work constitutes a milestone towards the practical achievement of this new class of MCs harnessing their advantages.

## Results

**Sample structure and STXM measurements** Figure 1 shows a scanning electron microscopy (SEM) image of the sample (Fig. 1a) as well as STXM images of out-of-plane core magnetizations (Fig. 1b) and in-plane curling magnetizations (Fig. 1c) in each of the five Py disks (see Methods for the sample dimensions). Here, the polarization and chirality configurations of the chain are $p$ = [+1,-1, +1, -1, +1] and $C$ = [-1, -1, -1, -1, +1], respectively (see Fig. 1d), as obtained from the STXM images, where $p$ = +1(-1) corresponds to upward (downward) core magnetization, and $C$ = +1 to counter-clockwise (CCW) and $C$ = -1 to clockwise (CW) in-plane curling magnetization. Note that the sample has the opposite core orientations between the NN disks.

In order to trigger an excitation of vortex gyration in the first disk, we launch a current pulse of 1.8 ns duration into the electrode stripline, resulting in a field pulse of about 2.4 mT strength [see the corresponding inset of Fig. 1(a)]. The propagation of vortex gyration excited at the first disk is driven by dipolar interaction between the NN disks where individual cores are shifted from their static center positions, thereby yielding a non-zero effective in-plane magnetization. Oscillatory motions of the individual cores are measured by STXM operated in the pump-and-probe sampling mode, which allows for imaging of the cores' out-of-plane magnetizations utilizing element-specific X-ray magnetic circular dichroism (XMCD) as magnetic contrast at a lateral resolution of about 25 nm and a temporal resolution as low as 35 ps (for further details, see Methods).

**Vortex-core gyration propagation along dipolar-coupled disks** Figure 2 shows the $x$ (red color) and $y$ (blue) components (Fig. 2a) of the displacements of the individual cores and their trajectories (Fig. 2b) in the disk plane, as measured by time-resolved STXM (see also Supplementary Movie 1). The experimental results (top of 2a and 2b) are compared with the corresponding micromagnetic simulations (bottom of 2a and 2b) performed using the OOMMF code (version 1.2a4)[44]. The characteristic beating patterns along with their modulation envelopes are observed in each of the five disks (Fig. 2a). Owing to the direct excitation of the first disk, a large-amplitude gyration in that disk is observed, and is then allowed to propagate towards the NN disk and beyond through the chain. The vortex-gyration transfer to the next disk and its further propagation are evidenced by the increase of the gyration amplitude in the second and remaining disks along with the concomitant and remarkable decrease of the first disk's gyration amplitude. The ratio between the maximum displacements in disk 5 and disk 1 is about 0.24. Since our pump-and-probe measurements are carried out within a time period of 60.8 ns, and the intrinsic damping of Py is not negligible but rather significant (as strong as $\alpha \sim 0.01$), we cannot clearly observe backward propagation bounced at the last (5$^{th}$) disk. However, the signature of weak reflection is evident by the increase of the gyration amplitude in the 4$^{th}$ disk at around 55 ns, as compared with the simulation result.

It has been reported that coupled gyrations in two-dipolar-coupled disks can be described by the superposition of the two normal modes[30,33,34,36,37]. Dipolar interaction between NN disks breaks the radial symmetry of the potential energy of each core, which depends on the disk pair's relative vortex-state configuration (both the polarization and chirality ordering). Analogously, for the case of the five-disk system used in this study, the beating patterns is the result of linear combinations of the five normal modes of coupled vortex gyration in the entire chain [for more information, see Supplementary Information B].

**Discrete wave modes of collective vortex gyration** To illustrate the collective vortex-gyration modes excited in the real sample, in Fig. 3(a) we plot the frequency spectra (red circles) of core motions in the individual disks as obtained from fast Fourier transformation (FFT) of the core position vector $\mathbf{X}_n$ multiplied by $C_n$ in the $n^{\text{th}}$ disk. We also compare the experimental results with micromagnetic simulation (green circles) for the five-disk model system and numerical calculation (blue circles) based on five coupled linearized Thiele equations[45] (see Supplementary Information A). Because of the intrinsic damping of core gyration in isolated disks, the peaks are broadened and overlapped with neighboring peaks to an extent, that they cannot be separated. Further deviations between the experimental data and micromagnetic simulations as well as numerical calculations can be attributed to sample imperfections and the chosen time steps (400 ps) between the snapshot images taken by STXM. Specifically, with regard to disks 2 and 4, only one peak of wide width appears in the experimental data whereas two peaks appear in the simulations and numerical data. In contrast, two clear peaks and a very weak third peak appear in disks 1 and 3, which is in quantitative agreement with the micromagnetic simulation and numerical calculation.

In order to clarify the presence of fundamental discrete modes, we conduct further numerical calculation on a five-disk model of the same dimensions and material parameters as those in the real sample, but with zero damping. The right panel of Fig. 3a shows the characteristic frequency spectrum of each disk. From disk 1 through disk 5, different peaks of contrasting FFT powers are observed. All of the five distinct peaks marked by $\omega_i$ (where $i$ = 1, 2, 3, 4, 5) are shown in disks 1 and 5. By contrast, the $\omega_3$ peak disappears in disks 2 and 4, while the $\omega_2$ and $\omega_5$ peaks disappear in disk 3. Each of the peaks of all of the modes is located at the same position in all of the disks.

From the inverse FFT of all of the peaks of each mode, we can extract the spatial correlations of core motions in the individual disks for each mode $\omega_i$. Figure 3(b) shows the trajectories of the orbiting cores in motion in the individual disks along with the profiles of the $C_nY_n$ component of the core positions in the five-disk chain. For all of the modes $\omega_i$, the individual core' gyration amplitudes are markedly distinct among the disks and modes. More interestingly, the collective motions of the individual cores in the whole chain represent certain wave forms of different wavelengths. The gyration amplitudes for all of the modes are symmetric with respect to the center of the chain, and are also completely pinned at imaginary disks at both ends, denoted disk 0 and disk 6 for the case of $N$ = 5. These features represent a standing-wave form of a certain wavelength in terms of collective vortex-gyration motions, being quite analogous to a string, the ends of which are attached to the left and right walls respectively, thus having no displacement. Accordingly, we can interpret the collective and discrete wave modes as in the no-damping case, based on the fixed boundary condition in such a 1D chain of finite disk number $N$. In this case, the boundary condition is given as $\psi|_{x=0} = \psi|_{x=(N+1)d_{\text{int}}} = 0$, where $\psi$ denotes the displacement of beads, $N$ is the number of elements, and $d_{\text{int}}$ is the inter-distance between the elements. From this boundary condition, the wave vectors of the allowed modes can be expressed simply as $k = m \cdot \pi / [(N+1)d_{\text{int}}]$, where $m = 1, 2, \ldots N-1$, and $N$ (For more information, see Supplementary Information B). Thus, the discrete (quantized) five modes' wave numbers of the collective vortex gyrations in the five-disk chain are coincident with the values of $k_m = m\pi/6d_{int}$, where $m=1, 2, 3, 4, 5$.

**Dispersion relation in coupled five-disk chain** As described above, collective vortex-gyration modes represent standing waves of discrete wavelengths (i.e., quantized $k$ values). Here, to extract the dispersion ($\omega$ - $k$ relation) of all of the modes, we perform FFTs of the collective core profiles for the individual modes according to $k_m = m\pi/6d_{int}$ (where $m=1, 2, 3, 4, 5$) with a fixed value of $d_{int}$=2250 nm for the real sample. Figure 4 shows the FFT powers in the $\omega$-$k$ spectra obtained from the experimental data, micromagnetic simulation and numerical calculation (for both cases of $\alpha$=0.01 and $\alpha$=0 in the numerical calculation). FFTs of each of the $X_n$ and $Y_n$, multiplied by $C_n$ (i.e., $C_n \cdot X_n$ and $C_n \cdot Y_n$), are performed. Using such reduced parameters of $C_nX_n$ and $C_nY_n$, we can consider only the polarity ordering for comparison between experimental and numerical calculation data (see Supplementary Information A).

The overall shape of the dispersion from the experimental data qualitatively agrees well with those from the micromagnetic simulations and numerical calculation, though they show quantitative discrepancy in the frequency and FFT power between each mode. As already mentioned above, discrepancies might be associated with the chosen measurement parameters, sample imperfections as well as a difference in the saturation magnetization between the experiment and micromagnetic simulations. The white solid lines indicate the result of the numerical calculation of an analytically derived explicit form for a 1D infinite chain (for the calculation, see Supplementary Information C). As noted earlier, the intrinsic damping of vortex-core gyration in isolated disks causes the broadening of the $\omega$ values (see

Fig. 4). For the case of no-damping, five discrete quantized modes without the $\omega$–value broadening are distinctly shown in the spectra (right panel).

Next, note that the overall shape of dispersion is concave down, that is, a higher frequency at $k=0$ and a lower frequency at $k=\pi/d_{int}$ for the case of $C_nX_n$, and concave up (vice versa) for $C_nY_n$. This reversal between $C_nX_n$ and $C_nY_n$ can be understood in terms of the lattice-number-dependent phase difference between the $x$ and $y$ components of the vortex-core positions. Since the gyration's rotational sense is determined by the polarization $p$ of a given disk, the phase difference between the $x$ and $y$ components of the core position vector in the $n^{th}$ disk is given as the product of $\pi/2$ and $p_n$. Accordingly, for the case of the antiparallel polarization between the NN disks as in the sample, the phase difference between the $x$ and $y$ components can be expressed as $\mathbf{R} \cdot \left(\frac{\pi}{d_{int}}\hat{\mathbf{k}}\right) - \frac{\pi}{2}$, where $\mathbf{R} = nd_{int}\hat{\mathbf{x}}$. This results in the shift of the $k$-vector in reciprocal space, as $k' = k - \pi/d_{int}$. Considering the real value of $\pi/d_{int} = 1.3963$ $\mu m^{-1}$ for $d_{int} = 2250$ nm), the experimental data are fully consistent for the $k$ shift by $\pi/d_{int}$ between $C_nX_n$ and $C_nY_n$, as shown in Fig. 4.

**Extension to semi-infinite or infinite 1D magnonic crystals** Based on the above approach, we can extend to a chain system comprised of a semi-infinite or infinite 1D chain composed of periodically arranged disks (referred to as 1D MCs). Specifically, we accomplish this by numerical calculation of a large number of disks (here, $N=201$) and an analytically derived dispersion equation for infinite chains. Here we also consider specific parallel and antiparallel ordering of both the $p$ and $C$ configurations between the NN disks: Type I: $[p_n, C_n] = [(-1)^{n+1}, 1]$ for the antiparallel $p$ and parallel $C$ ordering; Type II: $[(-1)^{n+1}, (-1)^{n+1}]$ for the antiparallel $p$ and $C$ ordering; Type III: $[1,1]$ for the parallel $p$ and $C$ ordering, and Type IV: $[1, (-1)^{n+1}]$ for the parallel $p$ and antiparallel $C$ ordering. Considering those additional degrees of freedom for both the *p* and *C* ordering, we analytically derive an explicit dispersion relation based on linearized Thiele equations of coupled vortex-core motions, taking into account the potential energy modified by dipolar interaction between only NN disks[30,31]. Here, for simplicity, we assume 1D arrays of an infinite number of equal-dimension disks. For zero damping ($\alpha$=0), the dispersion relation can be written as $\omega^2 = \omega_0^2 \zeta_\parallel^2 \zeta_\perp^2$ with $\zeta_\parallel^2 = 1 + 2C_n C_{n+1}(\eta_\parallel/\kappa)\cos(kd_{int})$ and $\zeta_\perp^2 = 1 - 2C_n C_{n+1} p_n p_{n+1}(\eta_\perp/\kappa)\cos(kd_{int})$, where $\kappa$ is the stiffness coefficient of the potential energy for isolated disks. $\eta_\parallel$ and $\eta_\perp$ represent the interaction strength along the *x* (here *x* is the bonding axis) and *y* axes, respectively (for the detailed derivation procedure, see Supplementary Information C). $p_n p_{n+1}$ = +1(-1) and $C_n C_{n+1}$ = 1(-1) indicate parallel (antiparallel) *p* and *C* ordering, respectively, between the NN disks. In this case, the wave vector *k* has a continuous value due to the infinite number of existing modes in such an infinite 1D chain. This explicit analytical form indicates that the dispersion relation is a function of an isolated disk's eigenfrequency $\omega_0$ and the coupling strength between the NN disks, that is, $\eta_\parallel$ and $\eta_\perp$, as well as those special *p* and *C* ordering.

The numerical calculation of the analytical form of $\omega^2(k) = \omega_0^2 \zeta_\parallel^2(k) \zeta_\perp^2(k)$ for four different types of vortex-state ordering noted above are displayed by the white lines in Fig. 5a, which are in excellent agreement with the dispersion spectrum from the FFTs of the $X_n$ components of the individual disks, which are obtained from the numerical calculation of *N* coupled Thiele equations for the *N*=201 system with damping ($\alpha$=0.01). While performing the FFTs, we imposed a periodic boundary condition to describe such a semi-infinite system in terms of traveling waves. Accordingly, the resultant *k*-values are given as $k = m \cdot (2\pi/Nd_{int})$, where *m* is any integer value under the constraint of $-\frac{\pi}{d_{int}} < k \leq \frac{\pi}{d_{int}}$. All of the dispersion curves are symmetric with respect to $k=0$, because the gyration is supposed to propagate from the center towards both ends.

We stress here that the overall shape of dispersion is determined by the sign of $p_n p_{n+1} C_n C_{n+1}$; such that concave up for $p_n p_{n+1} C_n C_{n+1}=1$, and concave down for $p_n p_{n+1} C_n C_{n+1} = -1$. Also, the band width is wider for the case of the antiparallel $p$ ordering than for the parallel $p$ ordering. The band-width variation in the $p$ ordering reflects the fact that the opposite polarization between NN disks has a stronger dipolar interaction (resulting in large frequency splitting) than does the same polarization, as noted in earlier reports [33-36]. This is caused by the rotational sense of stray fields around a given disk that is opposite to that of the core gyration. A rotating field efficiently couples into the circular eigenmode of gyration when the sense of rotation of the stray field coincides with the sense of gyration[46]. Consequently, the dipolar interaction between NN disks having the opposite polarization is stronger than that between those having the same polarization.

For a comprehensive understanding of dispersion variation according to the sign of $p_n p_{n+1} C_n C_{n+1}$, we extract the spatial distributions of collective vortex-core motions from the analytical derivation at specific values of $k=0$ (red lines) and $k = k_{BZ} = \pi/d_{int}$ (blue lines), as shown in Fig. 5(b). We calculate the dynamic dipolar interaction energy densities as a function of time for four different types of vortex-state ordering. The insets show the effective *in-plane* magnetizations of a given disk and both NN disks around it in a unit time period of $2\pi/\omega$. The rotating effective magnetizations $<\mathbf{M}_n>$ (gray-coloured wide arrows in each disk) in the NN disks and their relative orientations determine the characteristic dispersion that varies with both $p$ and $C$ orderings. The governing rule is determined by the dynamic dipolar interaction energy term: for the case where the NN disks' $<\mathbf{M}_n>$ is in parallel (antiparallel) orientation along the $x$ axis, their dipolar interaction energy is at the lowest

(highest) energy level, whereas for the case where their relative <**M**$_n$> is in parallel (antiparallel) orientation along the *y* axis, the energy is at the second highest (second lowest) energy level, as represented by the gray arrows in the three disks in Fig. 5b. Thus, the phase relation of NN disk's <**M**$_n$> is crucial to dynamic dipolar interaction, the overall value of which during a unit period is determined by $C_nC_{n+1}$ as well as $p_np_{n+1}$ (For more quantitative interpretation, see Supplementary Information D)

**Discussion**

We experimentally observed the wave modes of collective vortex-core gyration excitation along with their quantization and their dispersions in a chain of five coupled disks. With the help of an analytically derived explicit form, numerical calculation and micromagnetic simulation, those discrete modes can be well understood in terms of the relative orientations of rotating effective in-plane magnetizations and the dynamic dipolar interaction between the individual disks. Additional degrees of freedom of vortex-state ordering, including polarization and chirality, dramatically affect the phase relation of the dynamic dipolar interaction, thereby leading to changes in dispersion. Analogous to quantized lattice-vibration modes in solid crystals, the wave modes of vortex gyrations in periodically patterned magnetic dots are fundamental. This work enables the extension of coupled vortex disks to new types of MCs composed of ordered vortex-state disks, thus opening the way to control vortex-gyration propagations, band gaps, and widths of dispersions in 1D or 2D MCs.

Such new-type MCs might offer the advantages of limitless switchable-vortex-state and vortex-gyration-propagation endurance, low-power signal input through resonant excitation of vortex gyrations, and extremely low energy dissipation in information-processing devices when using negligible damping materials.

## Methods

**Sample preparation** The five Py disk chain is fabricated onto a 100-nm-thick silicon-nitride membrane using electron-beam lithography and lift-off techniques. Each disk has a thickness of 60 nm and a diameter of 2 μm. The center-to-center distance between neighboring disks is 2.25 μm. An 800-nm-wide Cu stripline of 120 nm thickness (with a gold cap of 5 nm thickness) is deposited onto the first disk [41].

**STXM measurement** Trajectories of the core motions of all five disks are directly observed using STXM by monitoring the out-of-plane core magnetizations at the MAXYMUS beamline (BESSY II; Helmholtz-Zentrum Berlin, Germany). The magnetic contrast is provided via XMCD at the Ni $L_3$ absorption edge (around 852.7 eV). The measurements showing the core polatization[47,48] are performed using negative circular-polarized x-rays (where an upward/downward core appears as a dark/white spot), whereas the measurements showing the chirality configuration are performed with the sample tilted 60° with respect to the beam axis, using positive circular-polarized x-rays (where a CW/CCW curling magnetization leads to a dark/bright contrast in the lower part of the disk). In the dynamic measurements, snapshot images of the individual core motions scanned in lateral steps of 8 nm are taken in time increments of 400 ps over a period of 60.8 ns after application of the field pulse at zero time

**Micromagnetic simulation** The Landau-Lifshitz-Gilbert (LLG) equation[49,50] of motion of local magnetizations is numerically solved for the model geometry identical to that of the sample applied in the experimental measurement, using the OOMMF code[44]. The material parameters corresponding to Py are as follows: saturation magnetization $M_s = 780 \times 10^3$ A/m, exchange stiffness constant $A_{ex} = 1.3 \times 10^{-11}$ J/m, and Gilbert damping constant $\alpha = 0.01$.

**Numerical calculation** Linearized coupled Thiele equations for vortex gyrations in $N$ coupled disks are numerically solved by taking into account the potential energy as modified by the dipolar coupling between the NN disks. In the numerical calculation, we use the numerical values of $\omega_0 = 2\pi \times 235\,\text{MHz}$, $G = 1.77 \times 10^{-12}\,\text{Js/m}^2$, $D = -4.66 \times 10^{-14}\,\text{Js/m}^2$, and $\kappa = 2.62 \times 10^{-3}\,\text{J/m}^2$, as obtained from the micromagnetic simulations performed on an isolated disk. The interaction strengths are determiend to be $\eta_\parallel = 9.36 \times 10^{-5}\,\text{J/m}^2$ and $\eta_\perp = 2.5 \times 10^{-4}\,\text{J/m}^2$ according to the relation between $\Delta\omega$ and the interaction strength coefficients[34], $\Delta\omega_{p_1 p_2} = \omega_0 (\eta_\perp - p_1 p_2 \eta_\parallel)/\kappa$, where we obtained $\Delta\omega = 2\pi \times 30\,\text{MHz}$ and $2\pi \times 15\,\text{MHz}$ for $p_1 p_2 = -1$ and $p_1 p_2 = 1$, respectively, from further micromagnetic simulations on a coupled two-disk system of the same dimensions as those of the real sample.

**Analytical derivation of dispersion for 1D infinite disk chain** We obtained dispersion relations of 1D infinite disk arrays for different $p$ and $C$ orderings noted in the text, based on the Thiele equation of motion of a single vortex core in isolated disks, but by taking into account the potential energy modified by dipolar interaction between only the NN disks (see Supplementary Information $C$ for further details.)

**Acknowledgements**

This work was supported by the Basic Science Research Program through the National Research Foundation of Korea (NRF) funded by the Ministry of Education, Science and Technology (Grant No. 20120000236). We acknowledge the support of Michael Bechtel, Eberhard Göring, and BESSY II, Helmholtz-Zentrum Berlin. Financial support from the Deutsche Forschungsgemeinschaft via the Sonderforschungsbereich 668 and the Graduiertenkolleg 1286 is gratefully acknowledged. This work has been also supported by the excellence cluster 'The Hamburg Centre for Ultrafast Imaging – Structure, Dynamics, and Control of Matter at the Atomic Scale' of the Deutsche Forschungsgemeinschaft. P. F. acknowledges support from the Director, Office of Science, Office of Basic Energy Sciences, Materials Sciences and Engineering Division, U.S. Department of Energy (contract no. DE-AC02-05-CH11231).

**Author contributions**

S.-K. K. and G. M independently led the project and conceived the main idea along with D.-S. H. and A. V., respectively. D.-S. H., K.-S. L. H. J., P. F., and S.-K. K. obtained preliminary experimental results of coupled vortex gyrations in different five-disk chain samples using magnetic transmission x-ray microscopy (MXTM) at the XM-1 beamline (at Advanced Light

## Additional information

**Supplementary Information** Additional information on analytical derivation, micromagnetic simulation and numerical calculation, and the time-resolved STXM movies is provided in Supplementary Information.

**Fig. 1**

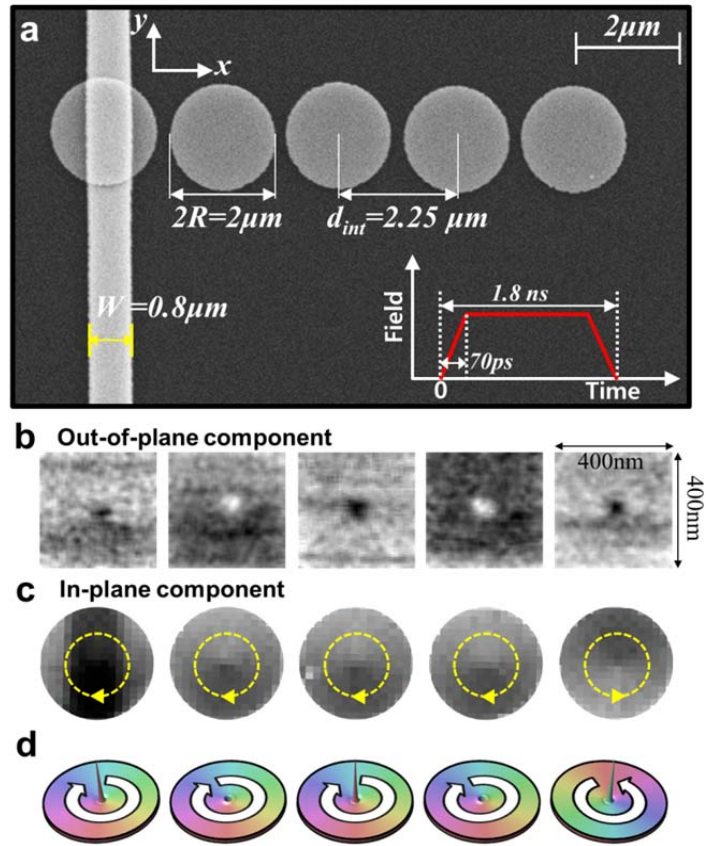

**Figure 1| SEM image and magnetization contrast for five-disk array**. **a**, SEM image of sample with a chain of five Py disks of identical dimensions and center-to-center distance and with a stripline for application of local magnetic field pulses to the left-end disk. The sample is deposited onto a silicon nitride membrane. The inset shows a schematic drawing of the field pulse used in the experiment. **b** and **c** represent initial vortex states in the five individual disks, out-of-plane magnetizations (the bright and dark spots correspond to the down and upward core orientations, respectively) and in-plane curling magnetizations (the curling orientations are indicated by the dashed arrows), as obtained from STXM measurements, respectively. **d** is a schematic illustration of the initial states of the sample.

**Fig. 2**

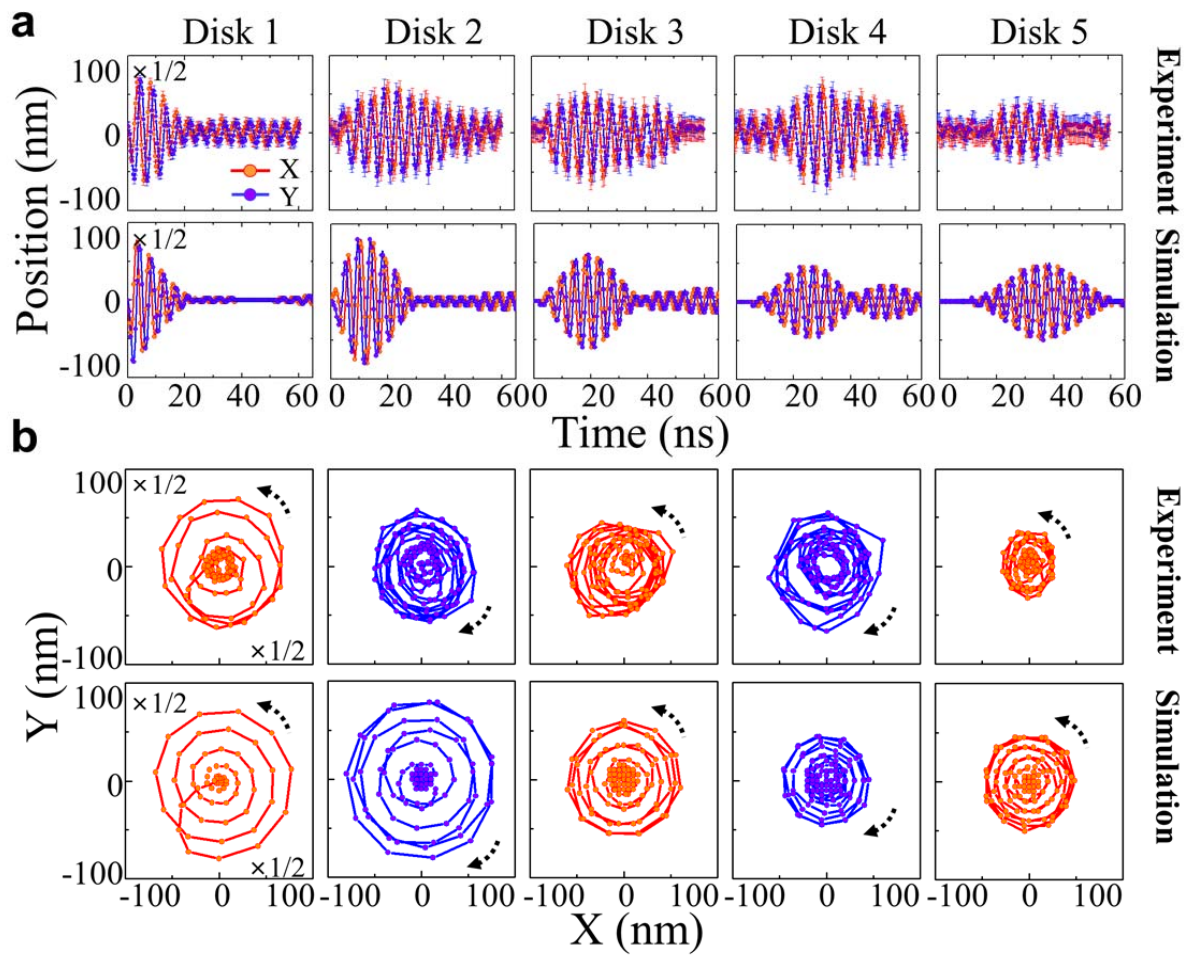

**Figure 2| Experimentally measured and simulated time-resolved trajectories of gyrating vortices.** **a**, Oscillatory *x* (red) and *y* (blue) components of vortex-core positions in individual disks as measured by STXM (upper row) and corresponding micromagnetic simulation data (bottom row). **b**, The trajectories of the vortex-core motion under a pulsed magnetic field during the time period $t$ = 0-60.8 ns. Dotted arrows indicate the sense of gyration of the individual cores.

**Fig. 3**

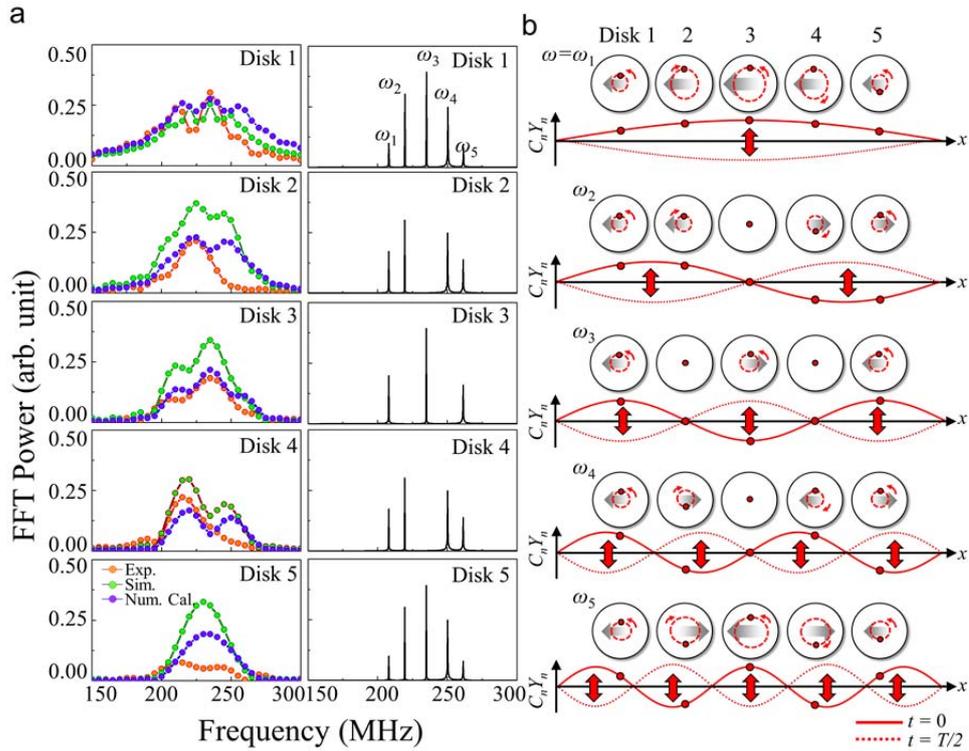

**Figure 3| Discrete vortex-gyration modes in a five-disk array. a**, Frequency spectra obtained from the experimental data, the micromagnetic simulation and the numerical calculation (FFTs of the numerically calculated oscillatory $C_n\mathbf{X}_n$ of the individual disks). The right column shows the numerically calculated frequency spectra for zero damping ($\alpha = 0$). **b**, Spatial distributions of the individual disks' core positions for the five discrete modes. The core trajectories are noted by the dashed lines inside the individual disks. Each dot on each trajectory represents the core position in the given disk. Just below each mode, the corresponding profile of the $C_nY_n$ components is indicated by the solid line (for $t = 0$) or dotted line ($t = T/2$), where $T = 2\pi/\omega$ is a time period for one cycle of gyration. In all of the FFTs, we applied the zero-padding technique to obtain 5 MHz resolution, except for the numerical calculation for $\alpha = 0$.

**Fig. 4**

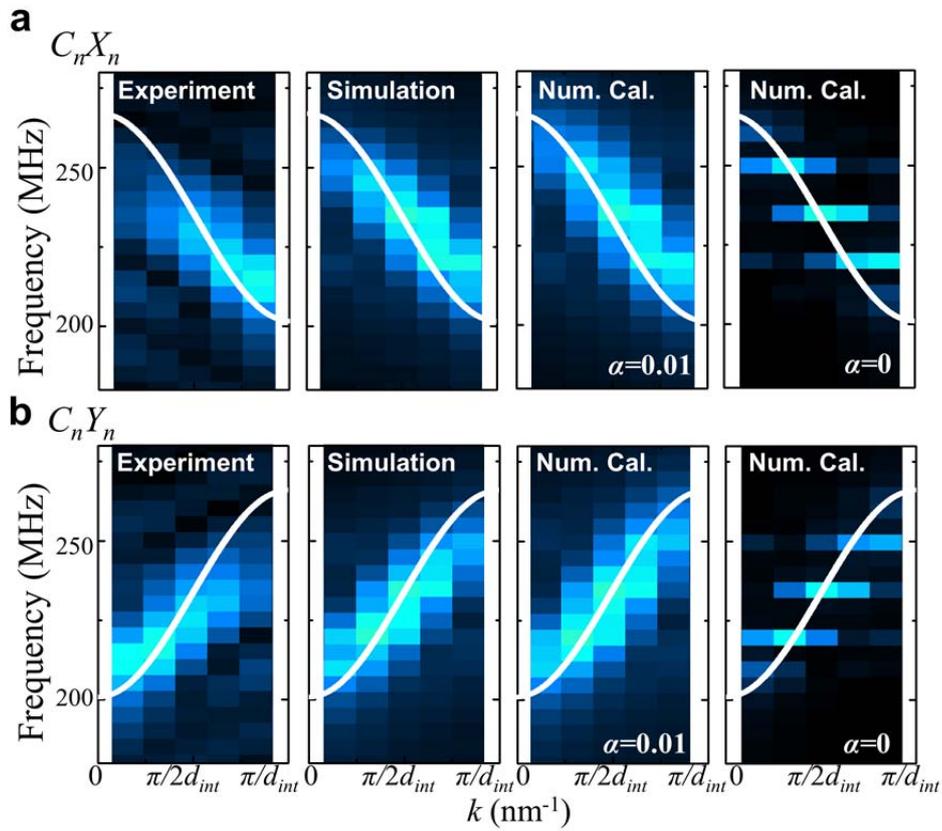

**Figure 4| Dispersion relations of collective vortex-gyration modes in a chain of five Py disks.** Dispersion relations for all excited collective modes, as extracted from FFTs of coupled oscillations of the vortex-core position vector $\mathbf{X}_n$ multiplied by $C_n$, i.e. **(a)** $C_n \cdot X_n$ **and (b)** $C_n \cdot Y_n$, obtained from experimental data, micromagnetic simulations, and numerical calculations with damping ($\alpha = 0.01$) and without damping ($\alpha = 0$). The white line indicates the analytically obtained dispersion curve for a 1D infinite array of the same dimensions and interdistance as in the simulations.

**Fig. 5**

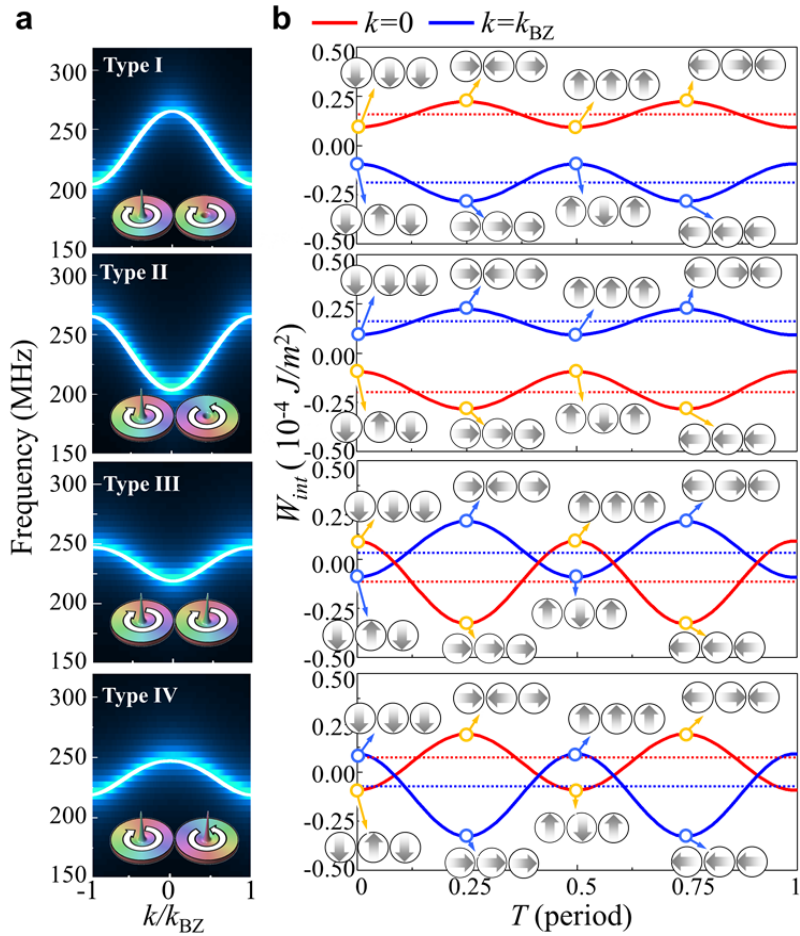

**Figure 5| Dispersion curves for parallel and antiparallel ordering of polarization $p$ and chirality $C$ as well as characteristic dynamic dipolar interaction energy density for $k=0$ and $k=k_{BZ}$ cases.** The four specific vortex-state orderings are: Type I for $[p_n, C_n]=[(-1)^{n+1}, 1]$, Type II for $[(-1)^{n+1}, (-1)^{n+1}]$, Type III for $[1, 1]$, and Type IV for $[1, (-1)^{n+1}]$. **a,** Dispersion relations of numerical calculation (blue-color spectra) for $N=201$ for which core gyration in the middle (101th disk) of the whole chain is displaced to ~ 200 nm, and of the numerical calculation (white thick lines) of the analytically derived equation for a 1D infinite chain. **b,** Dynamic dipolar interaction energy densities as a function of time for both $k=0$ and $k=k_{BZ}$, which are obtained from the analytical form of an infinite 1D array. The gray-colored wide arrows in each disk indicate the effective net magnetizations <$M_n$> induced by core shifts at

the given core positions. The size does not indicate the magnitude of the dynamic effective magnetizations. Specific <**M**$_n$> configurations are indicated by three disks and corresponding gray arrows.

SUPPLEMENTARY INFORMATION

Wave modes of collective vortex-gyration

in dipolar-coupled-dot-array magnonic crystals

Dong-Soo Han[1], Andreas Vogel[2], Hyunsung Jung[1], Ki-Suk Lee[1,†], Markus Weigand[3], Hermann Stoll[3], Gisela Schütz[3], Peter Fischer[4], Guido Meier[2] & Sang-Koog Kim[1,*]

1. National Creative Research Initiative Center for Spin Dynamics and Spin-Wave Devices, Nanospinics Laboratory, and Research Institute of Advanced Materials, Department of Materials Science and Engineering, Seoul National University, Seoul 151-744, Republic of Korea.

2. Institut für Angewandte Physik und Zentrum für Mikrostrukturforschung, Universität Hamburg, 20355 Hamburg, Germany.

3. Max-Planck-Institut für Intelligente Systeme, 70569 Stuttgart, Germany.

4. Center for X-ray Optics, Lawrence Berkeley National Laboratory, Berkeley CA 94720, USA.

*Correspondence and requests for materials should be addressed to S-K. Kim (email:

sangkoog@snu.ac.kr).

†Current address : School of Mechanical and Advanced Materials Engineering, Ulsan National Institute of Science and Technology, Korea

## A. Numerical calculation of *N* coupled Thiele equations for *N* coupled vortex gyrations

We derive linearized Thiele [S1,S2] equations for one-dimensional (1D) dipolar-coupled disks of finite disk number *N*, taking into account the potential energy modified by dipolar interaction[S3,S4] between only the next neighboring (NN) disks. The corresponding force-balance equation for the $n^{\text{th}}$ disk is given as

$$-\mathbf{G}_n \times \dot{\mathbf{X}}_n - \hat{D}_n \dot{\mathbf{X}}_n + \partial W / \partial \mathbf{X}_n = 0, \tag{S1}$$

where $\mathbf{X}_n = (X_n, Y_n)$ is the core position vector in the $n^{th}$ disk based on the collective coordinate ansatz. $\mathbf{G}_n = -Gp_n \hat{\mathbf{z}}$ is the gyrovector with its constant, $G = 2\pi L M_s / \gamma > 0$ (the saturation magnetization $M_s$, the gyromagnetic ratio $\gamma$, and the thickness of individual disk $L$), and $\hat{D}_n = D\hat{I}$ is the damping tensor with the identity matrix $\hat{I}$ and the damping constant $D$ [Ref. S5]. The $p_n$ represents the vortex core polarization in the $n^{th}$ disk. The total potential energy is given as $W = \sum_{n=1}^{N}\left(W_n(0) + \tfrac{1}{2}\kappa|\mathbf{X}_n|^2\right) + \sum_{n=2}^{N-1} W_{int}(\mathbf{X}_{n-1}, \mathbf{X}_n, \mathbf{X}_{n+1}) + \sum_{n=1}^{N} W_{\mathbf{H},n}$, where the first and second terms together are the potential energy with the stiffness coefficient $\kappa$ for isolated disks, and the third term $W_{int}$ is the dipolar interaction energy between only the NN disks. The last one, $W_{\mathbf{H},n} = -\mu_n (\hat{\mathbf{z}} \times \mathbf{H}_n) \cdot \mathbf{X}_n$, is the Zeeman energy term due to a driving force, where $\mu_n = \pi R L M_S \xi C_n$ (the radius of the individual disk $R$, the chirality of the vortex in the $n^{th}$ disk $C_n$, and $\xi = 2/3$) [Ref. S6]. The $W_{int}$, for a given $n^{th}$ disk, can be written as

$$W_{int} = C_{n-1}C_n \left(\eta_{\parallel} X_{n-1}X_n - \eta_{\perp} Y_{n-1}Y_n\right) + C_n C_{n+1}\left(\eta_{\parallel} X_n X_{n+1} - \eta_{\perp} Y_n Y_{n+1}\right), \quad \text{(S2a)}$$

for $n = 2, 3, \ldots N-1$,

$$W_{int} = C_n C_{n+1}\left(\eta_{\parallel} X_n X_{n+1} - \eta_{\perp} Y_n Y_{n+1}\right), \quad \text{(S2b)}$$

for $n = 1$

$$W_{int} = C_{n-1}C_n \left(\eta_{\parallel} X_{n-1}X_n - \eta_{\perp} Y_{n-1}Y_n\right), \quad \text{(S2c)}$$

for $n = N$. $\eta_{\parallel}$ and $\eta_{\perp}$ represent the interaction strength along the $x$ and $y$ axes (here $x$ is the bonding axis), respectively. The asymmetry between $\eta_{\parallel}$ and $\eta_{\perp}$ is due to the breaking of the radial symmetry of the potential energy of the isolated disks by their dipolar interaction. Finally, the coupled force-balance equations for the entire system are given as

$$-p_n G(dY_n/dt) - D(dX_n/dt) + \kappa X_n + \eta_{\parallel} C_n \left(C_{n+1} X_{n+1} + C_{n-1} X_{n-1}\right) = -\mu_n H_{y,n}, \quad \text{(S3a)}$$

$$p_n G(dX_n/dt) - D(dY_n/dt) + \kappa Y_n - \eta_\perp C_n (C_{n+1} Y_{n+1} + C_{n-1} Y_{n-1}) = \mu_n H_{x,n}, \quad \text{(S3b)}$$

where $n = 2, 3, \ldots N-1$,

$$-p_n G(dY_n/dt) - D(dX_n/dt) + \kappa X_n + \eta_\| C_n C_{n+1} X_{n+1} = -\mu_n H_{y,n}, \quad \text{(S4a)}$$

$$p_n G(dX_n/dt) - D(dY_n/dt) + \kappa Y_n - \eta_\perp C_n C_{n+1} Y_{n+1} = \mu_n H_{x,n}, \quad \text{(S4b)}$$

where $n = 1$, and

$$-p_n G(dY_n/dt) - D(dX_n/dt) + \kappa X_n + \eta_\| C_n C_{n-1} X_{n-1} = -\mu_n H_{y,n}, \quad \text{(S5a)}$$

$$p_n G(dX_n/dt) - D(dY_n/dt) + \kappa Y_n - \eta_\perp C_n C_{n-1} Y_{n-1} = \mu_n H_{x,n}, \quad \text{(S5b)}$$

where $n = N$. In the numerical calculation, we set the values of $H_{x,n}$ and $H_{y,n}$ to zero, except for the case of the $n^{th}$ disk, where an excitation of vortex gyration is triggered by an external field (i.e., $n=1$, when triggered at the left disk, $n=(N+1)/2$ (for odd $N$) or $N/2$ (for even $N$), when triggered at the center of the array).

Note that the chirality dependence of the governing equation can be eliminated by multiplying $C_n$ to the core position vector $X_n$ of a given $n^{th}$ disk. For an arbitrary $C$ configuration in a given chain, therefore, we can consider only the parallel and antiparallel $p$ ordering between the nearest-neighboring (NN) disks. In this case, $C_n \cdot X_n$ and $C_n \cdot Y_n$ correspond physically to the $-y$ and $x$ components of the in-plane net magnetization[S7], respectively, which is induced by the shift of the vortex core from its center position; Eqs. (S3-S5) can be regarded as the force-balance equations of the dipolar forces between the NN disks.

Using the above equations, for a certain value of $N$, we can numerically calculate coupled vortex gyrations of $N$ individual disks. The numerical values of $\omega_0 = 2\pi \times 235\,\text{MHz}$, $G=1.77 \times 10^{-12}$ Js/m$^2$, $\kappa = 2.62 \times 10^{-3}$ J/m$^2$, and $D = -4.66 \times 10^{-14}$ Js/m$^2$ are obtained from micromagnetic simulations performed on an isolated disk. We also obtain $\eta_\parallel = 9.36\times10^{-5}$ J/m$^2$ and $\eta_\perp = 2.56\times10^{-4}$ J/m$^2$ from further micromagnetic simulations performed on a coupled two-disk system of the same dimensions and interdistance as those used in this study, based on the relation[S8] of $\Delta\omega_{p_1p_2} = \omega_0(\eta_\perp - p_1p_2\eta_\parallel)/\kappa$, where $\Delta\omega = 2\pi\times30$MHz for $p_1p_2 = -1$ and $2\pi\times15$MHz for $p_1p_2 = 1$.

**B. Normal modes in a 1D array of finite disk number and fixed boundary condition**

In order to clarify the distinct modes of collective vortex-gyration excitation in a coupled disk chain of finite disk number $N$, we numerically calculate frequency spectra in the individual disks from fast Fourier transform (FFT) of vortex motions for the same dimensions and material parameters as those of the real sample, but with the zero-damping constant for three cases of $N=2$, 3 and 4, for example. As shown at the top of Fig. S1, for $N=2$, two distinct peaks appear in both disks, but for $N=3$, there are three peaks in disks 1 and 3 and only two peaks in disk 3. For $N = 4$, four peaks appear in all of the disks. These results clearly indicate that $N$ normal modes appear for a coupled $N$-disk chain, as reported in Refs [S3, S8-S13]. Each mode is labeled by $\omega_n$ with $n = 1, 2, ..., N$. For the individual modes in a given system of $N$, the spatial distributions of the individual cores (top) in collective motion and the corresponding displacement profiles of $C_nY_n$ (bottom, right) are plotted as obtained from the inverse FFTs of all of the peaks of all the disks for a given $\omega_n$ mode. The $C_nY_n$ component profiles (bottom, right) represent wave forms (red dotted or solid lines) of certain wave numbers. For the oscillatory motions of the $C_nY_n$ components of the individual cores, see Supplementary Movie 3. In accordance with the result for two dipolar-coupled vortex oscillators noted in earlier reports[S8-S13], the lower-frequency mode shows in-phase motion of the $C_nY_n$ component between disks 1 and 2, whereas the higher-frequency mode reveals anti-phase motion of the $C_nY_n$ component between disks 1 and 2. These two normal modes perfectly agree with standing waves of $k = \pi/(3d_{int})$ and $k = 2\pi/(3d_{int})$. For the case of $N=3$, there exists a standing-wave node (that is represented by no core motion in the middle disk) for the $\omega_2$ mode. This no-motion results in the absence of the $\omega_2$ mode's peak in the frequency spectra of disk 2. While, for the case of $N=4$, the disappearance of specific mode' peaks in the spectra does not happen. The reason is because any node is no longer located inside any disk of the entire chain. Note that the nodes can be observed only at the condition of $\sin(\mathbf{k} \cdot \mathbf{R}) = 0$, where $\mathbf{R} = nd_{int}\hat{\mathbf{x}}$ and $\mathbf{k} = m\pi/[(N+1)d_{int}]\hat{\mathbf{k}}$ with $m = 1, 2, \ldots N$. Thus, the nodes of standing waves are located inside the $n^{th}$ disk for the case where the constraint of $l\pi = nd_{int} \cdot [m\pi/\{(N+1)d_{int}\}]$ is satisfied, where $l$ is an integer.

Strikingly, those coupled vortex-gyration modes in the 1D chains of a given finite disk number are in excellent agreement with standing-wave forms of wave numbers, $k = m\pi/[(N+1)d_{int}]$, where $m = 1, 2, \ldots N$. In the numerical calculation of coupled vortex gyrations for a given $N$-disk system, although we do not assume a fixed boundary condition, all of the individual collective modes are completely pinned at both ends of the imaginary disks with the equal interdistance of $d_{int}$, (i.e., the $0^{th}$ and $N+1^{th}$ disk), so that we can adopt the fixed boundary condition of standing waves.

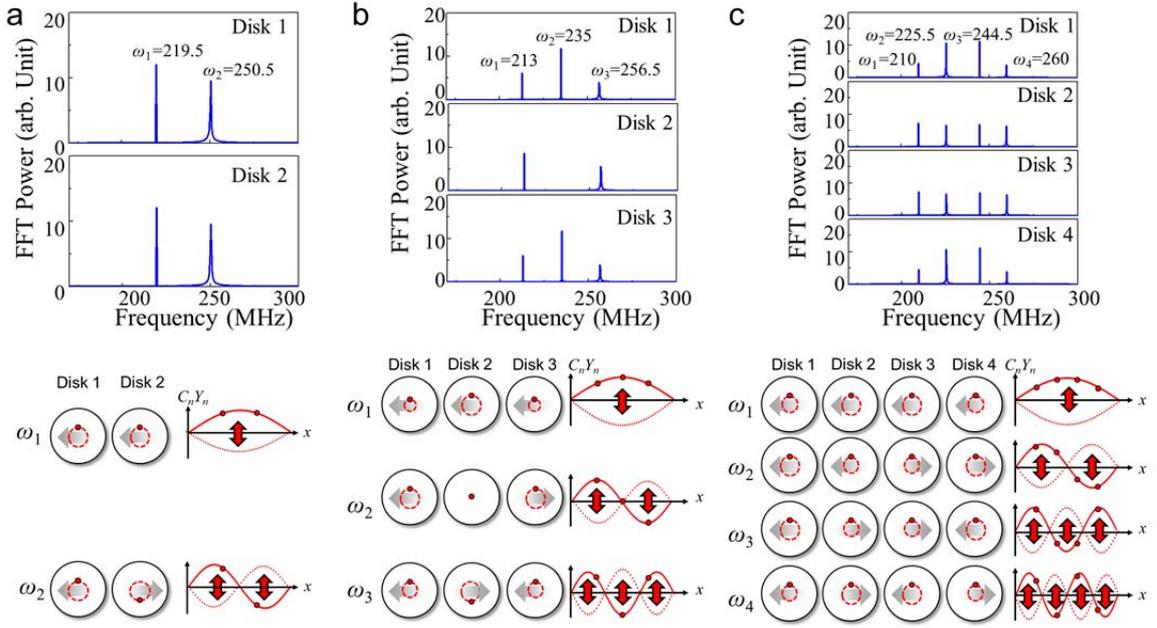

**Figure S1**. Frequency spectra and corresponding spatial distributions of distinct normal modes of dipolar-coupled magnetic disks of given disk numbers (**a**, $N=2$, **b**, $N=3$, and **c**, $N=4$). For all of the cases, the polarization ordering is indicated as $p_n = (-1)^{n+1}$ while holding counter-clockwise chirality ($C_n = +1$), where $n$ indicates the $n^{th}$ disk in a given chain. All the data are obtained from the FFTs of the $C_n\mathbf{X}_n$ components of the collective oscillatory motions of the individual disks calculated using coupled linearized Thiele equations shown in Supplementary Information **A**. The core trajectories are noted by the dashed lines inside the individual disks. Each dot on each trajectory represents the core position in the given disk. The wave profiles of the $C_nY_n$ components of the collective core motions are indicated by the red-colored solid ($t=0$) and dashed ($t=T/2$, where $T=2\pi/\omega$) lines.

## C. Analytical derivation of dispersion for 1D infinite disk array

We derived an explicit analytical form of dispersion for a 1D infinite disk array using the linearized Thiele equation[S1,S2] of motion, taking into account the potential energy modified by dipolar interaction[S3,S4] between only the NN disks of parallel and antiparallel polarization and chirality ordering. For simplicity, here, we assume 1D arrays of infinite disk number, and do not take into account Zeeman energy induced by external field (see Supplementary Information A). Then, the governing equation for the $n^{th}$ disk can be expressed as

$$-p_n G(dY_n/dt) - D(dX_n/dt) + \kappa X_n + \eta_\| C_n (C_{n+1} X_{n+1} + C_{n-1} X_{n-1}) = 0, \quad (S6a)$$

$$p_n G(dX_n/dt) - D(dY_n/dt) + \kappa Y_n - \eta_\perp C_n (C_{n+1} Y_{n+1} + C_{n-1} Y_{n-1}) = 0, \quad (S6b)$$

Considering the phase difference between $X_n$ and $Y_n$, the general solutions can be expressed as $X_n = \tilde{X} e^{i(ka_n - \omega t)}$ and $Y_n = \tilde{Y} e^{i(ka_n - \omega t + \pi p_n/2)}$ with $a_n = n d_{int}$, where $d_{int}$ is the center-to-center distance. Here $\tilde{X}$ and $\tilde{Y}$ are constants to be determined, whose ratio will specify the relative amplitude and phase of the gyrating effective magnetization within a unit cell. Incorporating these general solutions into Eqs. (S6a) and (S6b) yields

$$\omega_0 (1 + D^2/G^2) \left[ i\omega(D/G) / (\omega_0 (1 + D^2/G^2)) + 1 + C_n (C_{n+1} e^{ikd_{int}} + C_{n-1} e^{-ikd_{int}}) (\eta_\|/\kappa) \right] \tilde{X} - \omega \tilde{Y} = 0 \quad (S7a)$$

$$-\omega \tilde{X} + \omega_0 (1 + D^2/G^2) \left[ i\omega(D/G) / (\omega_0 (1 + D^2/G^2)) + 1 - C_n (C_{n+1} p_n p_{n+1} e^{ikd_{int}} + C_{n-1} p_n p_{n-1} e^{-ikd_{int}}) (\eta_\perp/\kappa) \right] \tilde{Y} = 0$$
(S7b)

with $\omega_0 \equiv \kappa G/(G^2 + D^2)$. According to the condition that the determinant of the 2×2 matrix is zero, the dispersion is given as

$$\omega^2 = \omega_0^2 (1 + D^2/G^2)^2 \left[ 1 + i\omega(D/G)/(\omega_0(1+D^2/G^2)) + C_n(C_{n+1} e^{ikd_{int}} + C_{n-1} e^{-ikd_{int}})(\eta_\|/\kappa) \right]$$
$$\times \left[ 1 + i\omega(D/G)/(\omega_0(1+D^2/G^2)) - C_n(C_{n+1} p_n p_{n+1} e^{ikd_{int}} + C_{n-1} p_n p_{n-1} e^{-ikd_{int}})(\eta_\perp/\kappa) \right] \quad (S8)$$

For the case of zero damping and for the given parallel and antiparallel $p$ and $C$ ordering,

Eq. (S8) can be rewritten as $\omega^2 = \omega_0^2 \zeta_\parallel^2 \zeta_\perp^2$ with $\zeta_\parallel^2 = 1 + 2C_n C_{n+1} (\eta_\parallel / \kappa) \cos(k d_{int})$ and $\zeta_\perp^2 = 1 - 2 C_n C_{n+1} p_n p_{n+1} (\eta_\perp / \kappa) \cos(k d_{int})$. The group velocity is then given as

$$\upsilon_g = \partial \omega / \partial k = d_{int} \omega_0 C_n C_{n+1} \left[ p_n p_{n+1} (\eta_\perp / \kappa)(\zeta_\parallel / \zeta_\perp) - (\eta_\parallel / \kappa)(\zeta_\perp / \zeta_\parallel) \right] \sin k d_{int} \quad (S9)$$

For all of the $p$ and $c$ orderings, the group velocity is zero at $k=0$ and $k_{BZ}$. All of the individual cores at $k = 0$ coherently move together, while for $k = k_{BZ}$ standing-wave forms result in nodes at every disk. This analytical form indicates that the group velocity can be manipulated by the selection of constituent materials, the dimensions of each disk, the separation distance as well as the $p$ and $C$ ordering.

**D. Phase relation of the net in-plane magnetizations between the NN disks**

To understand the variation of dynamic dipolar interaction between the NN vortices with the $p$ and $C$ ordering, we qualitatively determine the phase difference between two magnetic dipoles, which is to say, the net in-plane magnetizations <**M**> of two NN disks. The <**M**$_n$> of a given $n^{th}$ disk is induced by a shift of the vortex core from its center position. From earlier work[S7], it can be simply expressed, in terms of chirality $C$ and the vortex-core's position vector $\mathbf{X} \equiv X\hat{\mathbf{x}} + Y\hat{\mathbf{y}}$, as

$$\mathbf{m} \equiv V \cdot \langle \mathbf{M} \rangle_V \equiv \zeta C (Y\hat{\mathbf{x}} - X\hat{\mathbf{y}}), \quad (S10)$$

where $X$ and $Y$ represent the $x$ and $y$ components of the core displacement, respectively. Then, the effective magnetization in the $n^{th}$ disks for a certain mode of $\omega$ has the form

$$\mathbf{m}_n = \zeta R C_n (\hat{\mathbf{x}} \sin p_n \omega t - \hat{\mathbf{y}} \cos p_n \omega t) \quad (S11)$$

where $p_n$ and $\omega$ denote the vortex core's polarization in the $n^{th}$ disk and the angular frequency for a given mode, respectively. $\zeta$ is the proportionality constant, which is determined by the analytical model of vortex structure, and $R \equiv \sqrt{X^2 + Y^2}$. For simplicity, we set the initial phase of the $n^{th}$ disk to zero so that the initial orientation of $<\mathbf{M}_n>$ at $t=0$ is in the $-y$ (for $C_n=+1$) direction, except for $R=0$.

From the general solutions shown in Supplementary Information A, since the phase delay of the vortex-core position between the NN disks is given as $kd_{int}$ for a certain $k$ value, the effective magnetization vector in the $n+1^{th}$ disk takes the form of

$$\mathbf{m}_{n+1} = \zeta R C_{n+1}[\hat{\mathbf{x}} \sin(p_n \omega t + kd_{int}) - \hat{\mathbf{y}} \cos(p_n \omega t + kd_{int})] \quad . (S12)$$

The phase difference $\vartheta$ between $\mathbf{m}_n$ and $\mathbf{m}_{n+1}$ can be obtained by the inner product of their unit vectors, $\hat{\mathbf{m}}_n$ and $\hat{\mathbf{m}}_{n+1}$, and thus the phase difference can be expressed in terms of the polarization and chirality ordering between the NN disks as well as the angular frequency and wave vector of a given mode $\hat{\mathbf{m}}_n \cdot \hat{\mathbf{m}}_{n+1} = \cos \vartheta = C_n C_{n+1} \cos\left(kd_{int} - (p_{n+1} - p_n)\omega t\right)$.

Thus,

$$\vartheta = \vartheta_0 - (p_{n+1} - p_n)\omega t \quad (S13)$$

with $\vartheta_0 = kd_{int} + \pi\left(1 - \frac{1}{2} C_n C_{n+1}\right)$. For example, at $k=0$ ($k=k_{BZ}$), the initial phase difference, $\vartheta_0$, between the neighboring disks' $<\mathbf{M}>$ can be simply expressed as $\vartheta_0 = \pi - \frac{\pi}{2} C_n C_{n+1}$ ($\vartheta_0 = -\frac{\pi}{2} C_n C_{n+1}$). Therefore, for $k=0$, if $C_n C_{n+1}=1$ (-1), the initial phase difference between the neighboring disks' $<\mathbf{M}>$ has parallel (antiparallel) orientation along the $y$ axis. In the case of parallel polarization ordering, since their relative phase is always the same, the neighboring disks' $<\mathbf{M}_n>$ are in parallel (antiparallel) orientation along both the x and y axes, and thus the average dynamic dipolar energy is at the next-lowest (next-highest) level. On the other hand, in the case of antiparallel polarization ordering, the neighboring disks' $<\mathbf{M}_n>$, since their relative phase varies with time, are in parallel (antiparallel) and antiparallel (parallel) orientation along the y and x axes, respectively, and thus the average dynamic dipolar interaction energy is the highest (lowest), as clearly seen in Fig. 5b in the manuscript.

For $k=\pi/d_{int}$, if $C_nC_{n+1}=1$ (-1), the neighboring disks' $<\mathbf{M}_n>$ are in antiparallel and parallel orientation along the y and x axes, respectively, for antiparallel polarization ordering, whereas they are in antiparallel (parallel) orientation along the two axes for the parallel polarization ordering. Thus, for the case of the antiparallel polarization ordering, the average dynamic dipolar energy is at the lowest (highest) frequency level, while it is at the next-highest (next-lowest) frequency level for the case of parallel polarization ordering.

**E. Dispersion for different *N*-disk chains**

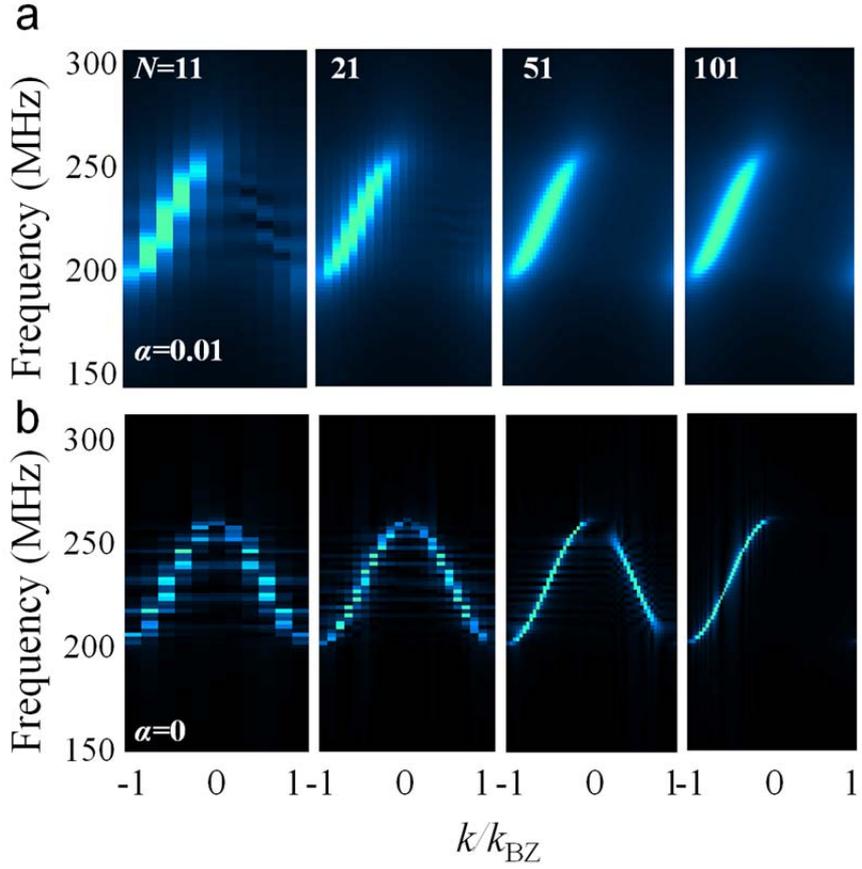

**Figure S2**. Dispersion relations in the same chain, but for different values of $N$ = 11, 21, 51, and 101, as obtained from numerical calculations (**a**) with damping and (**b**) without damping for a case of $[p_n, C_n] = [(-1)^{n+1}, 1]$. All of the curves are extracted from the FFTs of the numerically calculated vortex-core position vectors $\mathbf{X}_n$, for which core gyration is excited in the left end (n=1) disk of the array.

We numerically calculate coupled vortex gyrations for different values of $N$=11, 21, 51, and 101, but with the same dimensions as those of the real sample, for both cases (a) with and (b) without damping. As $N$ increases, the number of excited modes increases according to the condition, $k = m \cdot (2\pi/Nd_{int})$, where $m$ = 1, 2, … $N$. All of the dispersion curves are nonsymmetric with respect to $k$=0, because gyrations excited from the left-end disk of a given chain propagate towards the right-end disk. Due to damping, the spectral power in the $k < 0$ region is much stronger than in the $k > 0$ region, where the group velocities are positive in $k <$

0. For smaller $N$, such as $N = 11$, the quantization is very distinct, and a certain reflection from the right-end disk appears, leading to observable negative group velocities in the $k > 0$ region. As $N$ increases, the mode gaps become smaller and reflections do not occur or are very weak, because the gyration propagation cannot reach the right-end disk. In contrast, for zero damping, the reflection of gyration propagations from the right-end disk results in comparable spectral powers with negative group velocities up to $N=51$. For $N=101$, since this numerical calculation is the case just for 500 ns duration, the vortex gyration cannot reach the right end of the array within the duration, and thus the reflection is not included in the data. Accordingly, there is no spectral power in $k > 0$.

Correspondence and requests for materials should be addressed to S-K. Kim (email: sangkoog@snu.ac.kr).

## Supplementary References

[S1] Thiele, A. A. Steady-State Motion of Magnetic Domains. *Phys. Rev. Lett.* **30**, 230 (1973).

[S2] Huber, D. L. Dynamics of spin vortices in two-dimensional planar magnets. *Phys. Rev. B* **26**, 3758 (1982).

[S3] Shibata, J., Shigeto, K. & Otani Y. Dynamics of magnetostatically coupled vortices in magnetic nanodisks. *Phys. Rev. B* **67**, 224404 (2003).

[S4] Shibata, J. & Otani, Y. Magnetic vortex dynamics in a two-dimensional square lattice of ferromagnetic nanodisks *Phys. Rev. B* **70**, 012404 (2004).

[S5] Guslienko, K. Y. Low-frequency vortex dynamic susceptibility and relaxation in mesoscopic ferromagnetic dots *Appl. Phys. Lett* **89**, 022510 (2006).